\documentclass[12pt,twoside,headsepline,openright]{article}
\usepackage[english]{babel}
\usepackage{graphicx}

\begin{document}

 {\noindent \Large \sffamily \bfseries Entanglement of Orbital Angular \\
Momentum States of Photons}

\vspace{8mm}

 {\noindent Alois Mair,\footnote{Present address:
 Harvard-Smithsonian Center for Astrophysics
 60 Garden Street, Cambridge, MA 02138, USA}
 Alipasha Vaziri, Gregor Weihs, and Anton Zeilinger}

\vspace{8mm}

 {\noindent \itshape Institut f\"{u}r Experimentalphysik, Universit\"{a}t Wien \\
  \itshape Boltzmanngasse 5, 1090 Wien, Austria}

\vspace{8mm}

\noindent \textbf{Entanglement contains one of the most
interesting features of quantum mechanics, often named quantum
non-locality\cite{Schrodinger35a,Schrodinger35c}. This means
entangled states are not separable regardless of the spatial
separation of their components. Measurement results on one
particle of a two-particle entangled state define the state of
the other particle instantaneously with neither particle enjoying
its own well-defined state before the measurement.}

\textbf{So far experimental confirmation of entanglement has been
restricted to qubits, i.e. two-state quantum systems including
recent realization of three- \cite{Bouwmeester98a,Pan00a} and
four-qubit \cite{Sackett00a,Pan01a} entanglements. Yet, an ever
increasing body of theoretical work calls for entanglement in
quantum system of higher
dimensions\cite{Divincenzo00a,Bartlett00a}. For photons one is
restricted to qubits as long as the entanglement is realized
using the photon큦 polarization. Here we report the first
realization of entanglement exploiting the orbital angular
momentum of photons, which are states of the electromagnetic
field with phase singularities (doughnut modes). This opens up a
practical approach to multi-dimensional entanglement where the
entangled states do not only consist of two orthogonal states but
of many of them. We expect such states to be of importance for the
current efforts in the field of quantum computation and quantum
communication. For example, quantum cryptography with higher
alphabets could enable one to increase the information flux
through the communication channels
 \cite{Bechmann00a,Bechmann00b,Bourennane01a}.}

Multi-dimensional entanglement is another possibility, besides
creating multi-particle entanglement, for extending the usual
two-dimensional two-particle state. Thus far there also have been
suggestions\cite{multiport1,multiport2} and only a
proof-of-principle experiment\cite{multiport3} for realizing
higher order entanglement via multiport beam splitters. In the
following we present an experiment in which we employed a
property of photons namely the spatial modes of the
electromagnetic field carrying orbital angular momentum to create
multi dimensional entanglement. The advantage of using these
modes to create entanglement is that they can be used to define
an infinitely dimensional discrete (because of the quantization
of angular momentum) Hilbert space.

The experimental realization proceeded in the following two
steps, also reflected in the organization of the present paper.
First we confirmed that spontaneous parametric down-conversion
conserves the orbital angular momentum of photons. This was done
for pump beams carrying orbital angular momenta of $- \hbar$, 0,
and $+ \hbar$ per photon respectively. In a further step it was
shown that the state of the down-converted photons can not be
explained by assuming classical correlation in the sense that the
photon pairs produced are just a mixture of the combinations
allowed by angular momentum conservation. We proved that in
contrast they are a coherent superposition of these combinations
and hence they have to be considered as entangled in their orbital
angular momentum.

For paraxial light beams Laguerre-Gaussian ($\mathrm{LG}$) modes
define a possible set of basis vectors (Figure 1). As predicted by
Allen et al.\cite{Allen92a} and observed by He et al. \cite{He95a}
LG modes carry an orbital angular momentum for linearly polarized
light which is distinct from the angular momentum of photons
associated with their polarizations. This external angular
momentum of the photon states is the reason why they are often
have been suggested for gearing micro machines and it was shown
that they can be used as optical
tweezers\cite{Simpson97a,Galajda01a,Friese96a}.

To demonstrate the conservation of the orbital angular momentum
carried by the LG modes in spontaneous parametric down conversion
we investigated three different cases for pump photons possessing
orbital angular momenta of $- \hbar$, 0, and $+\hbar$ per photon
respectively. As a pump beam we used an Argon-ion laser at 351
$nm$ which we could operate either with a simple Gaussian mode
profile ($l = 0$) or in the first order LG modes $(l = \pm 1)$
after astigmatic mode conversion (for a description of this
technique see Ref.\cite{Beijersbergen93a}). Spontaneous parametric
down conversion was done in a 1.5 $mm$ thick BBO crystal cut for
type-I phase matching (that is both photons carry the same linear
polarization). The crystal cut was chosen such as to produce
down-converted photons at 702 $nm$ at an angle of $4^ \circ$ off
the pump direction.

The mode detection of the down-converted photons was performed
for Gaussian and LG modes. The Gaussian mode (l=0) was identified
using mono-mode fibers (Figure 2) in connection with avalanche
detectors. All other modes have a larger spatial extension and
therefore cannot be coupled into the mono-mode fiber. The LG modes
($l\neq 0$) were identified using mode detectors consisting of
computer generated holograms and mono-mode optical fibers (Figure
2).

 Computer generated holograms often have been exploited in the past for
creating LG modes of various orders.\cite{Artl98a}. Our holograms
were phase gratings 5 x 5 $mm^{2}$ in size with 20 lines per mm
which we first recorded on holographic films and bleached
afterwards to increase the transmission efficiency (Figure 2). We
made holograms which had one or two dislocations in the center and
designed them to have their maximum intensity in the first
diffraction order, so we could distinguish between LG modes $l=-2,
-1, 0, 1, 2 $ using all holograms in the first diffraction order
only for which order they have been blazed. For analyzing a LG
mode with a negative index the holograms were just rotated by
$180^ \circ$ around the axis perpendicular to the grating lines.
The total transmission efficiency of all our holograms was about
$80\%$ and they diffracted $18\%$ of the incoming beam into the
desired first order. These characteristics were measured at
632~nm as a laser source at 702~nm was not available to us.

The diffraction efficiency is not the only loss that occurs. Also,
we have to account for Fresnel losses at all optical surfaces
(95\% transmission), imperfect coupling into the optical fibers
(70\% for a Gaussian beam), non-ideal interference filters (75\%
center transmission), and the efficiency of the detectors (30\%).
A conservative estimate of all the losses yields an overall
collection efficiency of 2 to 3 percent. Comparing the
unnormalized ($l_{\mbox{pump}}=l_1=l_2=0$) coincidence rates of
about 2000~$s^-1$ to the singles count rates of about
100,000~$s^-1$ we deduce an efficiency of 2\%, well in agreement
with the above estimation.

The mode analysis was performed in coincidence for all cases
where mode filter 1 was prepared for analyzing LG modes $l_{1}=
0, 1, 2$ and mode filter 2 for those with $l_{2}= -2, -1, 0, 1,
2$. For analyzing a LG mode with mode index $l=0$, i.e. a
Gaussian mode, the dislocation of the hologram was shifted out of
the beam path. The beam was sent through the border of the
hologram where it acts as a customary grating without changing
the photon큦 angular momentum. The results are shown in Figure 3
for different values of orbital angular momenta of the pump beam.
Within experimental accuracy coincidences were only observed in
those cases where the sum of the orbital angular momenta of the
down converted photons was equal to the pump beam큦 orbital
angular momentum. However the absolute count rates of these cases
are not equal. This fact is most likely due to unequal emission
probabilities of the photons into the different modes in the down
conversion process.

These results confirm conservation of the orbital angular momentum
in parametric down-conversion. The achieved signal to noise ratios
were as high as $ V = 0.976 \pm 0.038$ and $ V = 0.916 \pm 0.009$
for pump beams with and without pump orbital angular momentum
respectively. $V$ is defined as $ V := \frac{I _{out} - I
_{in}}{I_{out} + I_{in}}$, where $I _{in}$ and $I _{out}$ denote
the maximum and the minimum of the coincidences with the
dislocation of the hologram in and out of the beam path
respectively.

 It is important to mention that only by using a
coincidence measurement we could show that the conservation of
the orbital angular momentum holds for each single photon pair.
In contrast, cumulative detection methods using many photons
result in an incoherent pattern \cite{Arlt99a} since each beam
from parametric down-conversion by itself is an incoherent
mixture. Therefore Arlt et al. \cite{Arlt99a} using these
classical detection methods which are in principle unsuitable at
the single photon level were led to believe that the orbital
angular momentum is not conserved in spontaneous parametric
down-conversion.

Given this experimental verification of the orbital angular
momentum conservation one may expect to find entanglement between
the two photons produced in the conversion process. But for
explaining the conservation of the orbital angular momentum the
photons do not necessarily have to be entangled. It would be
sufficient to assume classical correlation. However further
experimental results showed that the two-photon state goes beyond
classical correlation and indeed we were able to prove the
entanglement for photon states with phase singularities.

In order to confirm entanglement one has to demonstrate that the
two-photon state is not just a mixture but a coherent
superposition of product states of the various Gaussian and LG
modes which obey angular momentum conservation. For simplicity we
restricted ourselves to superpositions of two basis states only.
An important distinction between coherent superposition and
incoherent mixture of Gaussian and LG modes is that the latter
posses no phase singularity. This is because adding the spatial
intensity distributions of these two modes will yield a finite
intensity everywhere in the resulting pattern. In contrast, in a
coherent superposition the amplitudes are added and therefore the
phase singularity must remain and is displaced to an eccentric
location (Figure~4). It will appear at that location where the
amplitudes of the two modes are equal with opposite phase.
Therefore the radial distance of the singularity from the beam
center is a measure of the amplitude ratio of the Gaussian to the
LG components whereas the angular position of the singularity is
determined by their relative phase. Intuitively speaking the
position of the dislocation with respect to the beam is
equivalent to the orientation of a polarizer.

As discussed in Figure~2 such superpositions of LG and Gaussian
modes can experimentally be realized by shifting the dislocation
of the hologram out of the center of the beam by a certain small
amount. Hence in order to detect a photon having an orbital
angular momentum which is a superposition of the Gaussian and the
LG mode the hologram was placed in a position such that the
dislocation was slightly displaced from the beam center. In the
intensity pattern these modes possess an eccentric singularity
(Fig.~4). For demonstrating the entanglement we therefore shifted
one of the holograms and scanned the Gaussian mode filter on the
other side while recording the coincidences.

The results shown in Fig.~4 clearly verify the correlation in
superposition bases of the LG (l=$\pm 2$) and Gaussian (l=0)
modes. A closer analysis shows that there are two conditions
necessary to obtain the measured curves. First the shifted
hologram has to work as described above and second the source
must emit an angular momentum entangled state. Assume that the
source only emits classically correlated but not entangled
singularities. Then on the side with the shifted hologram the
various terms of the classical mixture would be projected onto a
state with displaced singularity leaving the total state again in
a mixture. Respecting the conservation of angular momentum we
would then have to sum the probabilities of the various
components on the other side resulting in a coincidence pattern
not containing any intensity zeroes. Such a coincidence pattern
would also be observed if a shifted hologram together with a
mono-mode detector would not be able to analyze for superposition
states.


An entangled state represents both correctly the correlation of
the eigenmodes and the correlations of their superpositions.
Having experimentally confirmed the quantum superposition for l=0
and l=$\pm 2$, it is reasonable to expect that quantum
superposition will also occur for the other states. Nevertheless,
ultimate confirmation of entanglement will be a Bell inequality
experiment generalized to more states \cite{Kaszlikowski00a}. Such
an experiment will be a major experimental challenge and it is in
preparation in our laboratory.

For a pump beam with zero angular momentum the emitted state must
then be represented by
\begin{equation}
\psi = C_{0,0}|0\rangle|0\rangle + C_{1,-1}|1\rangle|-1\rangle +
C_{-1,1}|-1\rangle|1\rangle + C_{2,-2}|2\rangle|-2\rangle +
C_{-2,2}|-2\rangle|2\rangle + ...... \label{zustand}
\end{equation}
since the LG modes form a infinite dimensional basis. Here the
numbers in the brackets represent the indices $l$ of the LG modes
and the $C_{i,j}$ denote the corresponding probability amplitude
for measuring $|i\rangle|j\rangle$. The state (\ref{zustand}) is
a multi-dimensional entangled state for two photons, which in
general will also contain terms with $\mathrm{p \neq 0}$. It means
neither photon in state (1) possesses a well-defined orbital
angular momentum after parametric down conversion. The
measurement of one photon defines its orbital angular momentum
state and projects the second one into the corresponding orbital
angular momentum state.

It is conceivable to extend these states to multi-dimensional
multi-particle entanglement in the future. A steadily increasing
body of theoretical work calls for entanglement of quantum
systems of higher dimensions \cite{Divincenzo00a,Bartlett00a}.
These states have applications in quantum cryptography with
higher alphabets and in quantum teleportation. Since such states
increase the flux of information it is conceivable that they will
be of importance for many other applications in quantum
communication and quantum information too. Also the possibility
to use these photon states for driving micro machines and their
application as optical tweezers make them versatile and
auspicious for future technologies
\cite{Simpson97a,Galajda01a,Friese96a}.

After completion of the experimental work presented here related
theoretical work was brought to our
attention.\cite{theory1,theory2}


\section*{Acknowledgements}

This work was supported by the Austrian Fonds zur F\"{o}rderung der
wissenschaftlichen Forschung (FWF).

Correspondence and request for materials should be addressed to
A.Z. (e-mail: anton.zeilinger@univie.ac.at)

\noindent \textbf{Captions:}

Figure~1: The wave front (top) and the intensity pattern (bottom)
of the simplest Laguerre Gauss (LG$_{p}^{l}$) or doughnut mode.
The index $l$ is referred to as the winding number and ($p+1$) is
the number of radial nodes. Here we only consider cases of $p =
0$. The customary Gaussian mode can be viewed as LG mode with
$l=0$. The handedness of the helical wave fronts of the LG modes
is linked to the sign of the index $l$ and can be chosen by
convention. The azimuthal phase term $ e^{il \phi}$ of the LG
modes results in helical wave fronts. The phase variation along a
closed path around the beam center is $ 2 \pi l$. Therefore in
order to fulfill the wave equation the intensity has to vanish in
the center of the beam.

Figure~2:   Experimental setup for single-photon mode detection.
After parametric down conversion each of the photons enters a
mode detector consisting of a computer generated hologram and a
mono-mode optical fiber. By diffraction at the hologram the
incoming mode undergoes a mode transformation in a way that a LG
mode can be transformed into a Gaussian mode. Since it has a
smaller spatial extension than all LG modes, only the Gaussian
mode can be coupled into the mono-mode fiber. Thus observation of
a click projects the mode incident on the fiber coupler into the
Gaussian mode. The hologram is a phase grating with $\Delta m$
dislocations in the center blazed for first order diffraction. An
incoming Gaussian laser beam passing through the dislocation of
the hologram is diffracted by the grating and the n-th
diffraction order becomes a LG mode with an index $l = n \Delta
m$ and vice versa. Intuitively speaking the phase dislocation
exerts a ``torque'' onto the diffracted beam because of the
difference of the local grating vectors in the upper and lower
parts of the grating. This ``torque'' depends on the diffraction
order n and on $\Delta m$. Consequently the right and left
diffraction orders gain different handedness. Reversing this
process a photon with angular momentum $\Delta m \hbar$ before
the grating can be detected by the mono-mode fiber detector
placed in the first diffraction order. A photon with zero angular
momentum (Gaussian mode) is detected by diffracting the beam at
the border of the hologram faraway from the dislocation. All our
measurements were performed in coincidence detection between the
two down-converted photons.

Figure~3: Conservation of the orbital angular momentum.

Coincidence mode detections for photon 1 and photon 2 in 15
possible combinations of orthogonal states were performed. This
was done for a pump beam having an orbital angular momentum of $-
\hbar$, 0, and $ +\hbar$ per photon respectively. Coincidences was
observed in all cases where the sum of the orbital angular
momenta of the down converted photons were equal to the pump
beams orbital angular momentum. The coincidence counts for each
fixed value of the orbital angular momentum of photon 1 was
normalized by the total number of coincidences varying the orbital
angular momentum of photon 2.

Figure~4:

Experimental evidence of entanglement of photon states with phase
singularities: The dislocation of the hologram in the beam of
photon 1 is shifted out of the beam center step by step (top,
middle, bottom). In these positions this hologram together with
the mono-mode fiber detector projects the state of photon 1 into a
coherent superposition of LG and Gaussian modes. The mode filter
for photon 2 with the hologram taken out makes a scan of the
second photon큦 mode in order to identify the location of its
singularity with respect to the beam center. The coincidences
show that the second photon is also detected in a superposition
of the LG and the Gaussian mode. Classical correlation would yield
a coincidence picture which is just a mixture of Gaussian and LG
modes. In that case the intensity minimum would remain in the
beam center but would become washed out. In the experiment a
hologram with two dislocations in the first diffraction order was
used. This results in a superposition of the l=0 and l=2 modes.

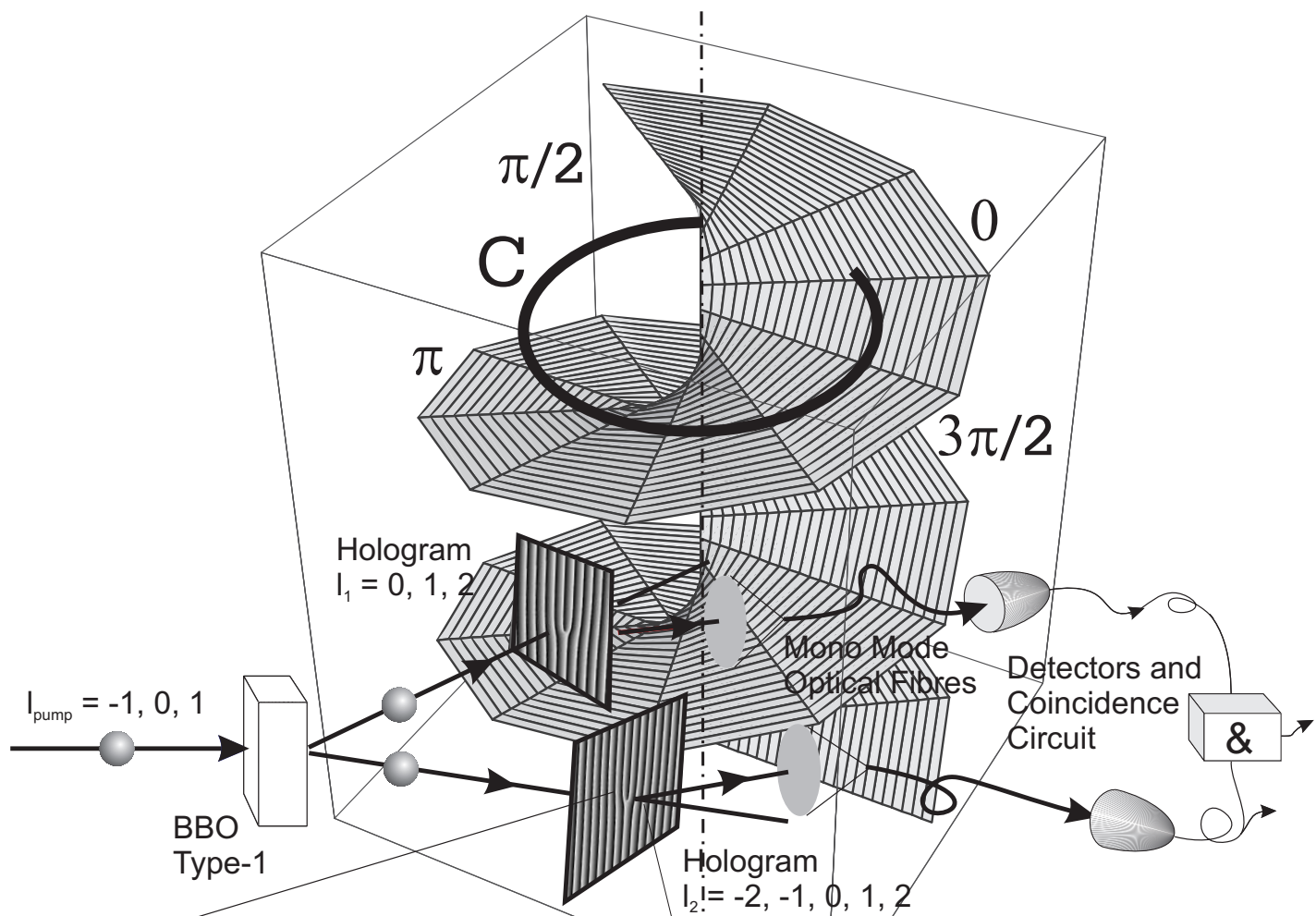
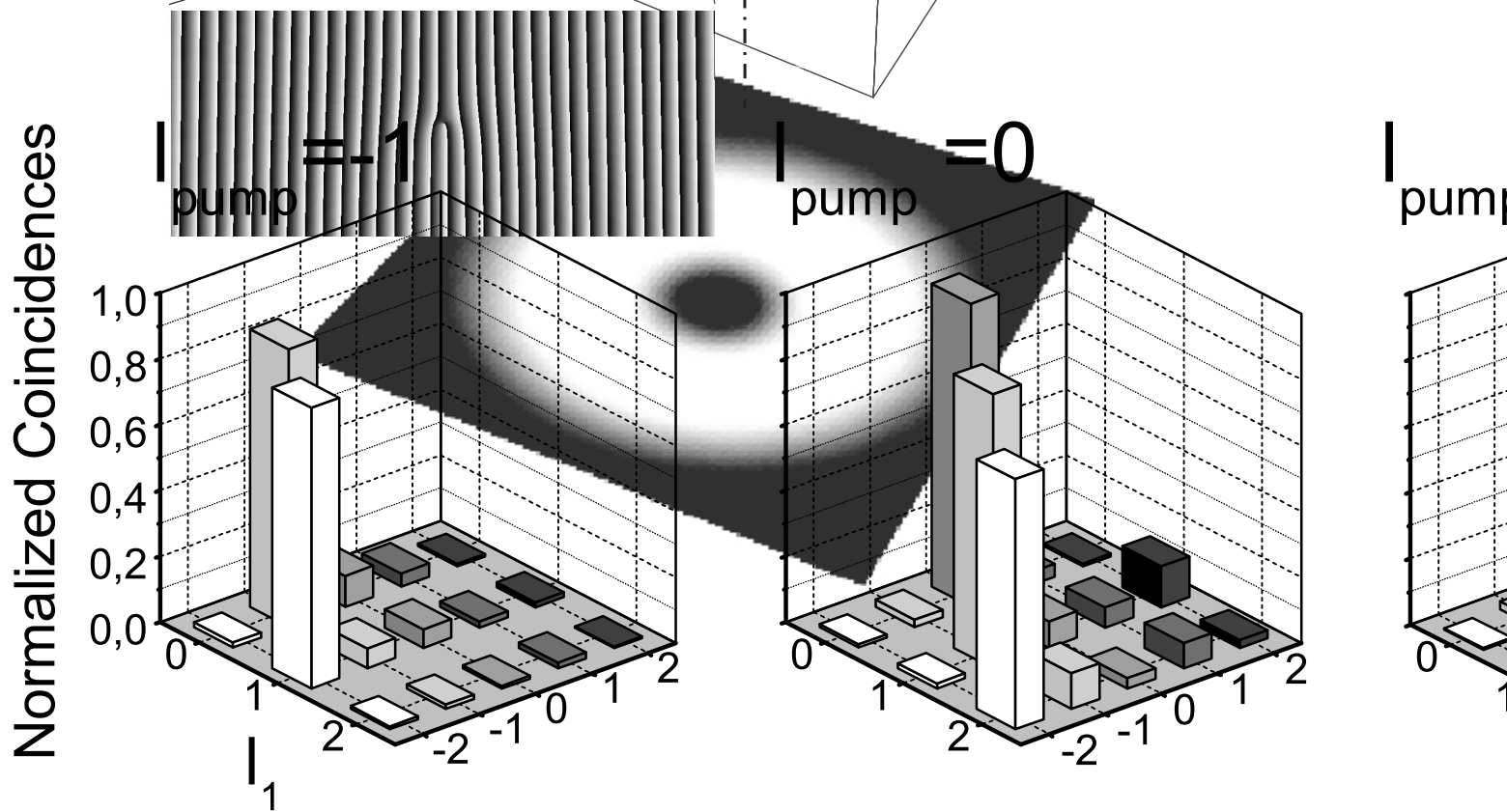
\end{document}